\documentclass[epj,referee]{svjour}

\usepackage{latexsym}
\usepackage{graphicx}%
\usepackage{dcolumn}
\usepackage{bm}

\begin{document}

\title{Chemical reactivity of ultracold polar molecules: investigation 
 of $\mbox{H}+\mbox{HCl}$ and $\mbox{H}+\mbox{DCl}$ collisions} 
\author{P. F. Weck\inst{}\thanks{\email{weckp@unlv.nevada.edu}} 
 \and N. Balakrishnan\inst{}\thanks{\email{naduvala@unlv.nevada.edu}} 
 }                     
%
%
\institute{Department of Chemistry, University of Nevada Las Vegas, 
 4505 Maryland Parkway, Las Vegas, NV 89154, USA
}
\date{Received: date / Revised version: date}
%
\abstract{
%
Quantum scattering calculations are reported for the 
$\mbox{H}+\mbox{HCl}(v,j=0)$ and $\mbox{H}+\mbox{DCl}(v,j=0)$ 
collisions for vibrational levels $v=0-2$~ of the diatoms. 
Calculations were performed for incident kinetic energies in the range 
$10^{-7}-10^{-1}~\mbox{eV}$, for total angular momentum $J=0$ 
and $s$-wave scattering in the entrance channel of the 
collisions.
Cross sections and rate coefficients are characterized by resonance 
structures due to quasibound states associated with the formation 
of the H$\cdots$HCl and H$\cdots$DCl van der 
Waals complexes in the incident channel. For the $\mbox{H}+\mbox{HCl}(v,j=0)$ 
collision for $v=1,2$, reactive scattering leading to $\mbox{H}_2$ formation is 
found to dominate over non-reactive vibrational quenching in the ultracold regime. 
Vibrational excitation of HCl from $v=0$ to $v=2$ increases the zero-temperature 
limiting rate coefficient by about 8 orders of magnitude.   
}

\authorrunning{Weck and Balakrishnan}
\titlerunning{Chemical reactivity of ultracold polar molecules}
\maketitle

\section{Introduction} \label{intro}

Over the last several years, much progress has been made in 
cooling, trapping, and manipulating  molecules at ultracold temperatures
\cite{wei98,bet00,joc03,gre03,zwi03,cub03} and 
Bose-Einstein condensation (BEC) of diatomic molecules has recently been 
demonstrated \cite{joc03,gre03,zwi03}. 
The experimental
breakthrough that led to the creation of molecular Bose-Einstein condensates 
starting from fermionic atoms 
provides unique
opportunities to
study the crossover regime between Bardeen-Cooper-Schrieffer-type
superfluidity of momentum pairs and BEC of molecules \cite{tim01,reg04,bar04,bou04}, 
a topic that has been of long interest to the high temperature 
superconductivity community. 

Collisional studies of ultracold molecules have received considerable attention
in recent years \cite{balak98,balak00,balak03,stoe03,til04,sold02,volpi03}
and the possibility of quantum collective effects in chemical reactions
involving ultracold molecules is of particular interest \cite{hei00,hop01}. 
Polar molecules are another class of molecules that have received
important attention in recent experiments.
The anisotropic, long-range character of the electric 
dipole-dipole interactions of polar molecules also designates 
them as potential candidates for scalable quantum computation 
schemes using electric dipole moment 
couplings \cite{bar95,bre99,pla99,dem02}. 
The techniques developed so far  for creating ultracold molecules
fall into three different categories, namely buffer-gas 
cooling of paramagnetic molecules \cite{wei98,doy95,dec99,fri99}, electrostatic cooling of polar 
molecules \cite{mad99,cro01,bet02,van02} and photoassociation of ultracold 
atoms \cite{fio98,tak98,nik00,gab00,pic04}. 
While the first two approaches have been successful in trapping polar molecules 
at cold temperatures of 10-100 mK, the 
creation of ultracold ($T \simeq 100~\mu$K) polar neutral ground state
  molecules (KRb) was 
achieved only recently by photoassociation in a magneto-optical trap 
\cite{mancini04}.
Formation of electronically excited RbCs molecules by 
photoassociation in a laser-cooled mixture of $^{85}$Rb and $^{133}$Cs 
atoms \cite{ker04,kerma04} has also been reported.

Photoassociation creates molecules in highly excited vibrational levels
and their lifetime is restricted by collisions leading to 
vibrational relaxation and/or chemical reactivity \cite{bal01,bod02}.
The effect of vibrational excitation on quenching rate coefficients in the 
ultracold regime has been explored before but similar studies on chemical 
reactivity have not been reported.
In this work, we report quantum scattering calculations 
of atom-diatom reactions at cold and ultracold temperatures in which 
the diatom is taken to be a highly polar molecule. 
Previous studies of ultracold chemical reactions investigated nonpolar molecules 
like H$_2$ \cite{bal01}, its isotopic counterparts \cite{bod02,balakri03} or 
alkali metal dimers \cite{sold02}.
Here, we investigate the bimolecular $\mbox{H}+\mbox{HCl}$ and 
$\mbox{H}+\mbox{DCl}$ collisions at 
low and ultralow energies for which the reaction proceeds mainly by quantum tunneling of 
the exchanged atom through a barrier along the reaction path.
As elementary steps in the $\mbox{H}_2+\mbox{Cl}_2$ reaction system, which plays 
a major role in chemical kinetics and in atmospheric chemistry,    
the gas-phase $\mbox{H}+\mbox{HCl}$ and $\mbox{Cl}+\mbox{H}_2$ reactions and 
their isotopic variants have received important attention both 
theoretically \cite{bra93,aoi99,aoi00,han01,che02,yao03,aoi95,man99,she01,sko02,bal03} 
and experimentally \cite{aoi00,ake89a,ake89b,bar91,bro98,taa99}. However, these 
kinetics studies and experiments have been carried out in the temperature range 
$195~\mbox{K} \leq T \leq 3020~\mbox{K}$ and no scattering calculations have been reported 
so far for these reactions in the cold and ultracold regimes, to our knowledge.     

In this paper, we present results for state-to-state and initial-state-selected 
probabilities and cross sections for both reactive and non-reactive channels 
of the $\mbox{H}+\mbox{HCl}(v,j=0)$ and $\mbox{H}+\mbox{DCl}(v,j=0)$ 
collisions for vibrational levels $v=0-2$~ in the ground electronic state. 
The presence of pronounced resonance structures due to quasibound states associated 
with van der Waals complexes in the initial channel is discussed.   
Finally, reaction rate coefficients for $\mbox{H}_2$ and HD formations 
are also presented as a function of the temperature and in the zero-temperature limit. 
We show that vibrational excitation of HCl and DCl dramatically increases the rate 
coefficients in the ultracold regime.

\section{Calculations}  \label{sec:1}

The quantum mechanical coupled-channel hyperspherical coordinate method of Skouteris 
et al. \cite{sko00} is 
used to solve the Schr\"{o}dinger equation for the motion of the three nuclei on the 
parametric representation of the single Born-Oppenheimer potential energy 
surface (PES) developed by Bian and Werner (BW) \cite{bia00}. 
The small effect of the fine-structure observed in similar reactions \cite{bal03,ale98,ale00} 
supports our choice to neglect the spin-orbit splitting in the Cl($^2P$) atom.      
Although the accuracy of this potential energy surface for ultracold collision studies is 
questionable, based on our experience we believe that major findings of our study will not 
be affected if a more accurate ClH$_2$ potential surface is used. 


Scattering calculations were performed for a total molecular angular momentum $J=0$ 
and $s$-wave scattering in the incident channel of the 
$\mbox{H}+\mbox{HCl}(v,j=0)$ and $\mbox{H}+\mbox{DCl}(v,j=0)$ collisions for 
vibrational states $v=0-2$.
We note that in the case of weak trapping potentials, which are 
expected to allow long decoherence times in 1D trap arrays of quantum 
computers \cite{dem02}, only $s-$wave scattering is expected to play a significant 
role \cite{kaj03}.
Because at very low kinetic and internal energies these reactions proceed mainly by 
quantum tunneling, the resulting reaction probabilities are very small and 
particular attention must be paid to convergence. Extensive convergence tests 
of the initial-state-selected and state-to-state reaction probabilities 
have been carried out, with respect to the number of rovibrational levels included 
in the basis set, $j_{max}$, the maximum value of the hyperradius, $\rho_{max}$, and the step 
size for the log derivative propagation, $\Delta{\rho}$.    
\begin{figure} 
\begin{center}
\resizebox{0.49\textwidth}{!}{\includegraphics{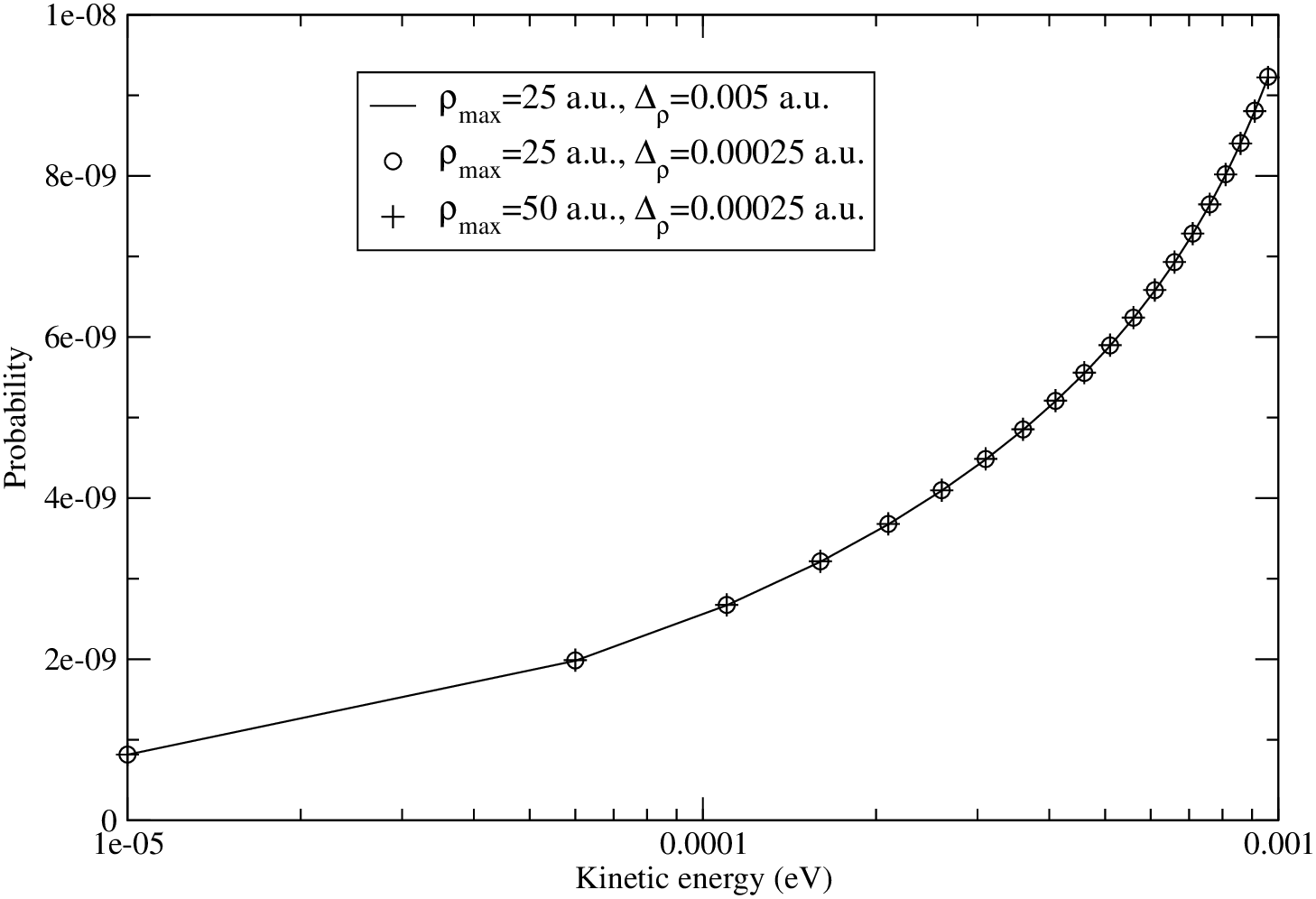}}
\end{center}
\caption{\label{fig:1} Reaction probability for H$_2$ formation in 
 $\mbox{H}+\mbox{HCl}(v=0,j=0)$ collisions as a function of the 
 incident kinetic energy for different values of $\rho_{max}$ and 
 $\Delta{\rho}$. 
 }
\end{figure}
Figure \ref{fig:1} shows the total reaction probability for H$_2$ formation in 
$\mbox{H}+\mbox{HCl}(v=0;j=0)$ collisions as a function of the incident kinetic 
energy for different values of $\rho_{max}$ and $\Delta{\rho}$. Convergence of 
the total reaction probability with an accuracy of the order of $10^{-10}$ was 
obtained with $\rho_{max}=25.0~\mbox{a.u.}$ and $\Delta{\rho}=0.005~\mbox{a.u.}$ 
for kinetic energies in the range $10^{-5}-10^{-3}~\mbox{eV}$. Using these 
values of $\rho_{max}$ and $\Delta{\rho}$, a more stringent convergence test 
consisted in the analysis of the product rotational ($j'$) distribution. For an 
incident kinetic energy of $10^{-3}~\mbox{eV}$, convergence of the state-to-state 
probability for $\mbox{H}+\mbox{HCl}(v=0,1;j=0)$ and 
$\mbox{H}+\mbox{DCl}(v=0,1;j=0)$ was achieved to within $10^{-10}$ using 
$j_{max}=15$ and a cutoff 
internal energy $E_{max}=2.9~\mbox{eV}$ in any channel. The resulting basis sets 
for HCl and DCl collisions consisted of 376 and 456 basis functions, respectively.   
\begin{figure} 
\begin{center}
\resizebox{0.49\textwidth}{!}{\includegraphics{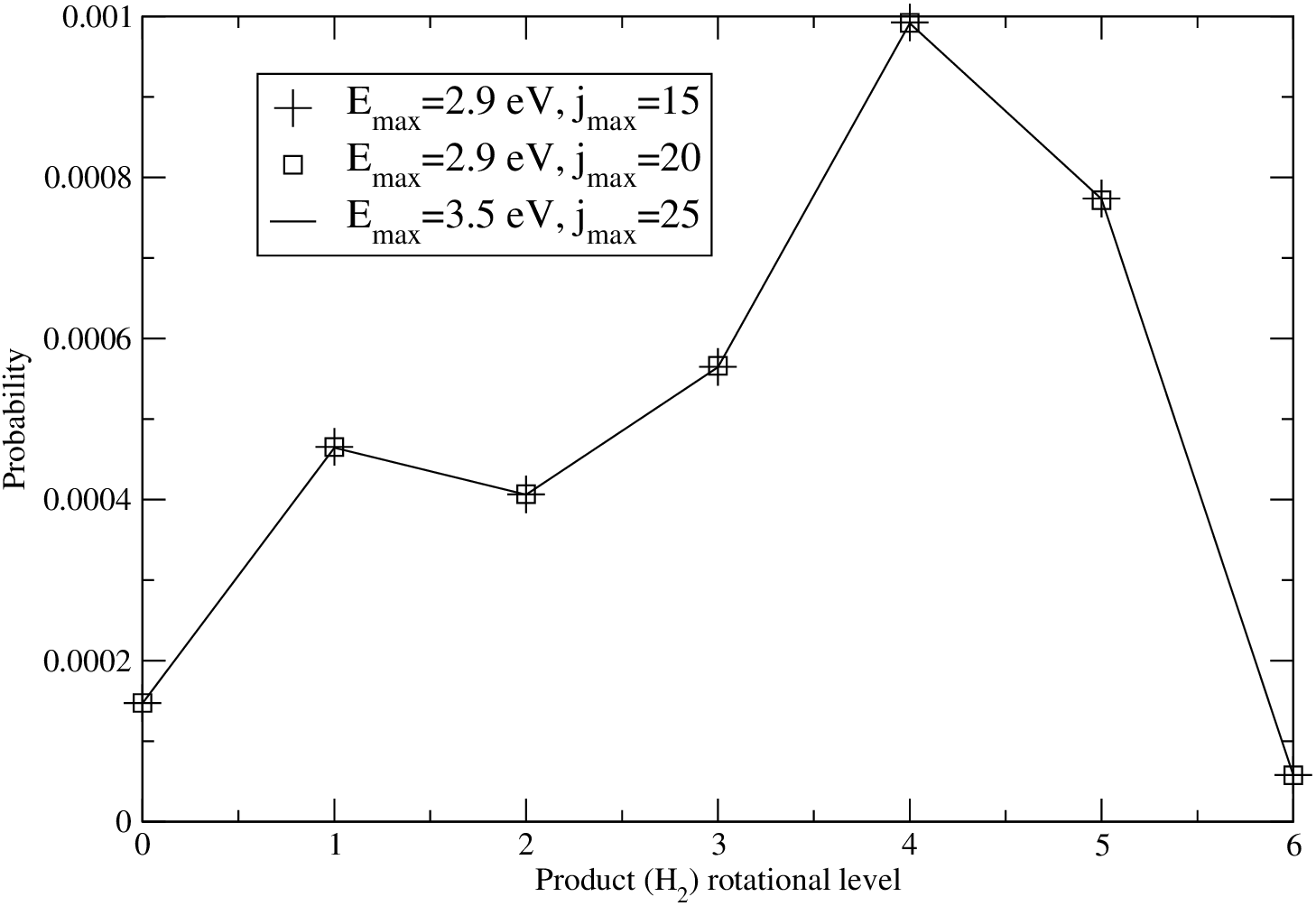}}
\end{center}
\caption{\label{fig:2} State-to-state reaction probability for 
 $\mbox{H}_2(v'=0,j')$ formation as a function of $j'$ in $\mbox{H}+\mbox{HCl}(v=1,j=0)$ collisions, 
 for different values of $E_{max}$ and $j_{max}$, at an incident kinetic energy 
 of $10^{-3}~\mbox{eV}$, $\rho_{max}=25.0~\mbox{a.u.}$, 
 and $\Delta{\rho}=0.005~\mbox{a.u.}$  
 }
\end{figure}
Figure \ref{fig:2} illustrates the convergence of the state-to-state reaction 
probability with respect to $j_{max}$ and $E_{max}$ for $\mbox{H}_2(v'=0,j')$ 
formation in $\mbox{H}+\mbox{HCl}(v=1,j=0)$ collisions, for an incident kinetic energy of 
$10^{-3}~\mbox{eV}$, $\rho_{max}=25.0~\mbox{a.u.}$, and $\Delta{\rho}=0.005~\mbox{a.u.}$ 
Convergence of the state-to-state probabilities for the
$\mbox{H}+\mbox{HCl}(v=2;j=0)$ and $\mbox{H}+\mbox{DCl}(v=2;j=0)$ collisions was achieved 
using larger basis sets of 721 and 891 basis functions, respectively, corresponding 
to $j_{max}=25$ and $E_{max}=3.5~\mbox{eV}$. On the basis of these convergence tests, values of 
$j_{max}=15$ and $E_{max}=2.9~\mbox{eV}$ for $v=0,1$ and of $j_{max}=25$ and $E_{max}=3.5~\mbox{eV}$ 
for $v=2$ were adopted for the calculations reported hereafter.

\section{Results and discussion}  \label{sec:2}

\begin{figure} 
\begin{center}
\resizebox{0.49\textwidth}{!}{\includegraphics{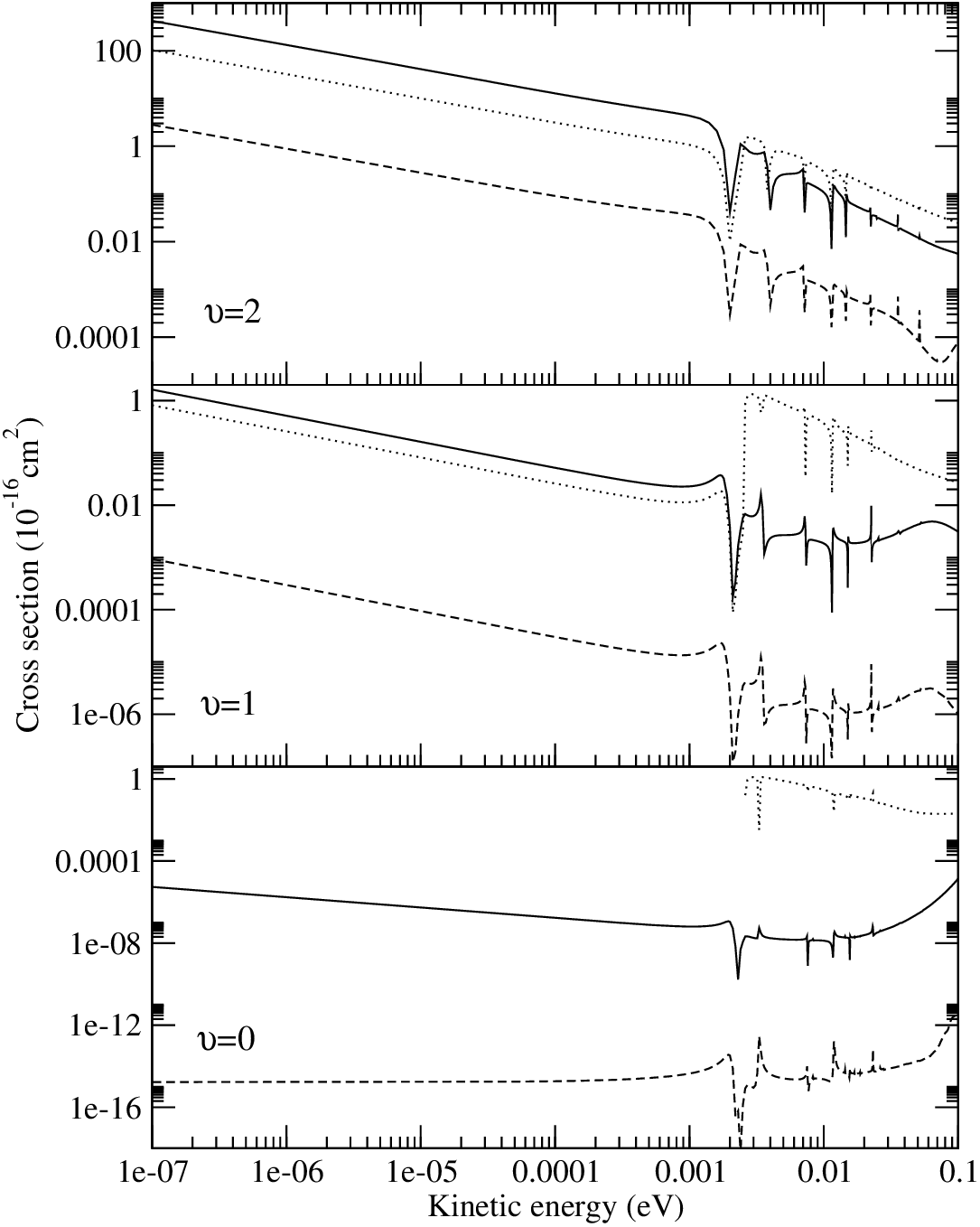}}
\end{center}
\caption{\label{fig:3} Cross sections for H$_2$ and HCl formation and nonreactive 
 scattering in $\mbox{H}+\mbox{HCl}(v,j=0)$ collisions, for $v=0-2$, as a function of the 
 incident kinetic energy. Dotted curve: nonreactive scattering; dashed curve: reactive 
 HCl product channel; solid curve: H$_2$ product channel.}
\end{figure}

The cross sections for H$_2$ and HCl formation and for nonreactive 
scattering in $\mbox{H}+\mbox{HCl}(v,j=0)$ collisions are shown in 
Figure \ref{fig:3}, for $v=0-2$, for incident kinetic energy in the range 
$10^{-7}-10^{-1}~\mbox{eV}$. For HCl molecules initially in their ground vibrational state, 
the reaction proceeds mainly by quantum tunneling through the barrier, yielding small 
values for the cross sections. 
We note that for the $v=0$, $j=0$, initial state nonreactive channels are open only for 
incident kinetic energy larger than $2.585\times 10^{-3}~\mbox{eV}$, corresponding to 
the energy value for 
rotational excitation to the first excited level $j=1$ of the ground vibrational state.
The sharp rise in the cross section for nonreactive scattering at energies above 
$2.585\times 10^{-3}~\mbox{eV}$ is due to rotational excitation to the $j=1$ level.
In the zero-temperature limit, the cross section for the abstraction reaction leading 
to H$_2$ formation is about 9 orders of magnitude larger than the exchange mechanism leading 
to HCl formation, consistent with the fact that the transmission 
coefficient for tunneling of the H atom through a finite potential barrier is larger than 
for heavier atoms like chlorine \cite{bala03}. As $v$ increases and the barrier height 
decreases, the $\mbox{H}_2/\mbox{HCl}$ product branching ratio decreases, with values differing by 3 orders 
of magnitude for $v=1$, and 2 orders of magnitude for $v=2$ at an incident kinetic energy 
of $10^{-7}~\mbox{eV}$.     
For energies below $10^{-4}~\mbox{eV}$, cross sections reach the Wigner regime \cite{wig48} 
where they vary inversely as the velocity and their ratios become constant. 
For the $v=0$, $j=0$ case, the hydrogen exchange process is indistinguishable from elastic 
scattering and the cross section attains a constant value in the Wigner limit as expected for 
elastic scattering. 
For kinetic energies larger 
than $10^{-3}~\mbox{eV}$, pronounced resonance structures appear in the cross sections due to quasibound 
states associated with the formation of the H$\cdots$HCl van der Waals complex in the 
initial channel, as reported previously for different molecular 
systems \cite{par93,alt98,lar00,wei03}. 
%
\begin{figure} 
\begin{center}
\resizebox{0.49\textwidth}{!}{\includegraphics{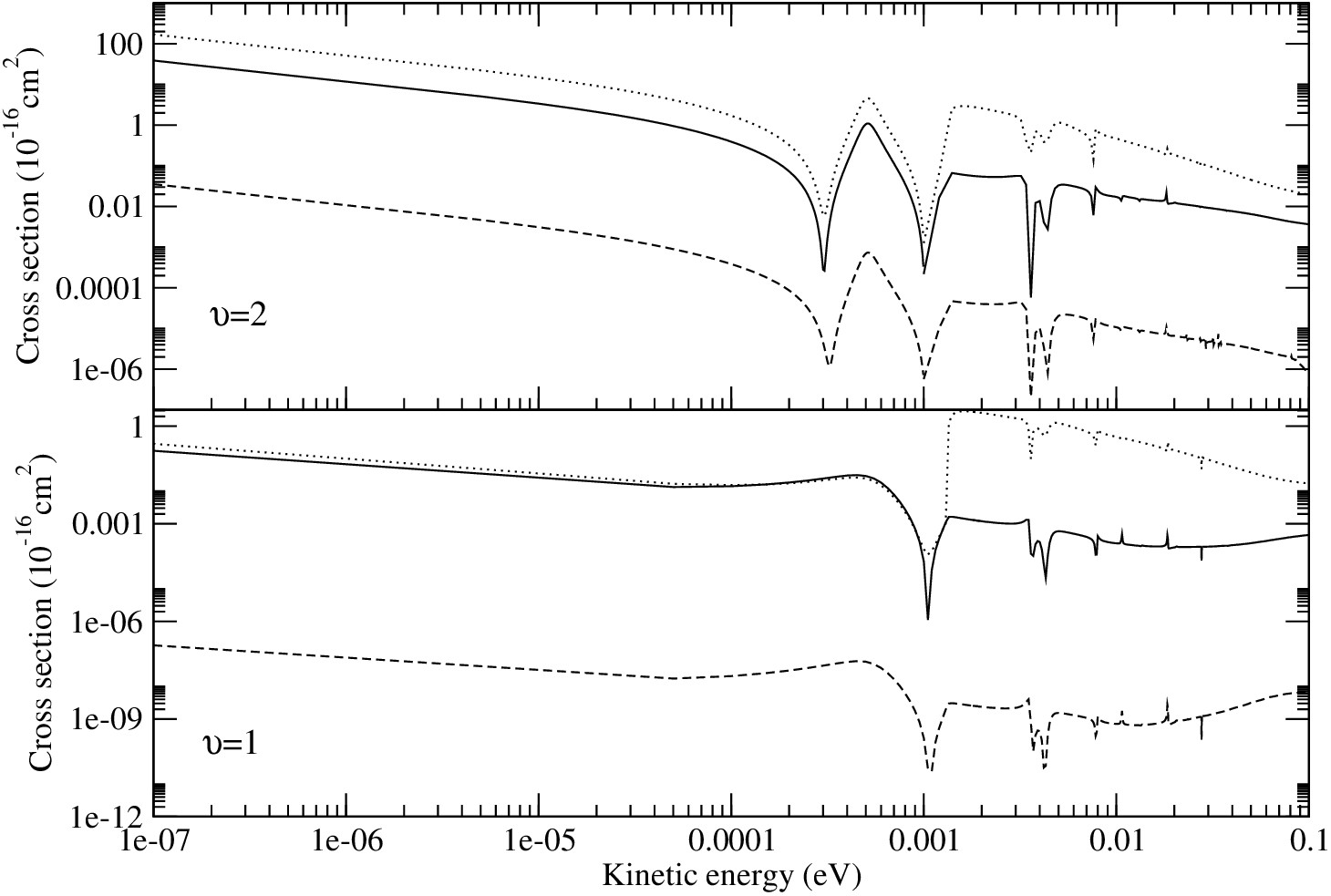}}
\end{center}
\caption{\label{fig:4} Cross sections for HD and HCl formation and nonreactive 
 scattering in $\mbox{H}+\mbox{DCl}(v,j=0)$ collisions, for $v=1,2$, as a function of the 
 incident kinetic energy. Dotted curve: nonreactive scattering; dashed curve: reactive 
 HCl product channel; solid curve: HD product channel.}
\end{figure}
Figure \ref{fig:4} shows cross sections for HD and HCl formation as well as 
nonreactive scattering in $\mbox{H}+\mbox{DCl}(v,j=0)$ collisions as functions of 
the incident kinetic energy. Cross sections are presented only for the first two 
vibrationally excited states $v=1$ and 2 due to the negligible values obtained for 
the $v=0$ level of DCl. 
While nonreactive cross sections have similar magnitude as for $\mbox{H}+\mbox{HCl}(v,j=0)$ 
collisions, reactive cross sections are several orders of magnitude smaller for the 
deuterated reaction, the difference being attributed to the less efficient tunneling of 
the heavier D atom. 

\begin{figure} 
\begin{center}
\resizebox{0.49\textwidth}{!}{\includegraphics{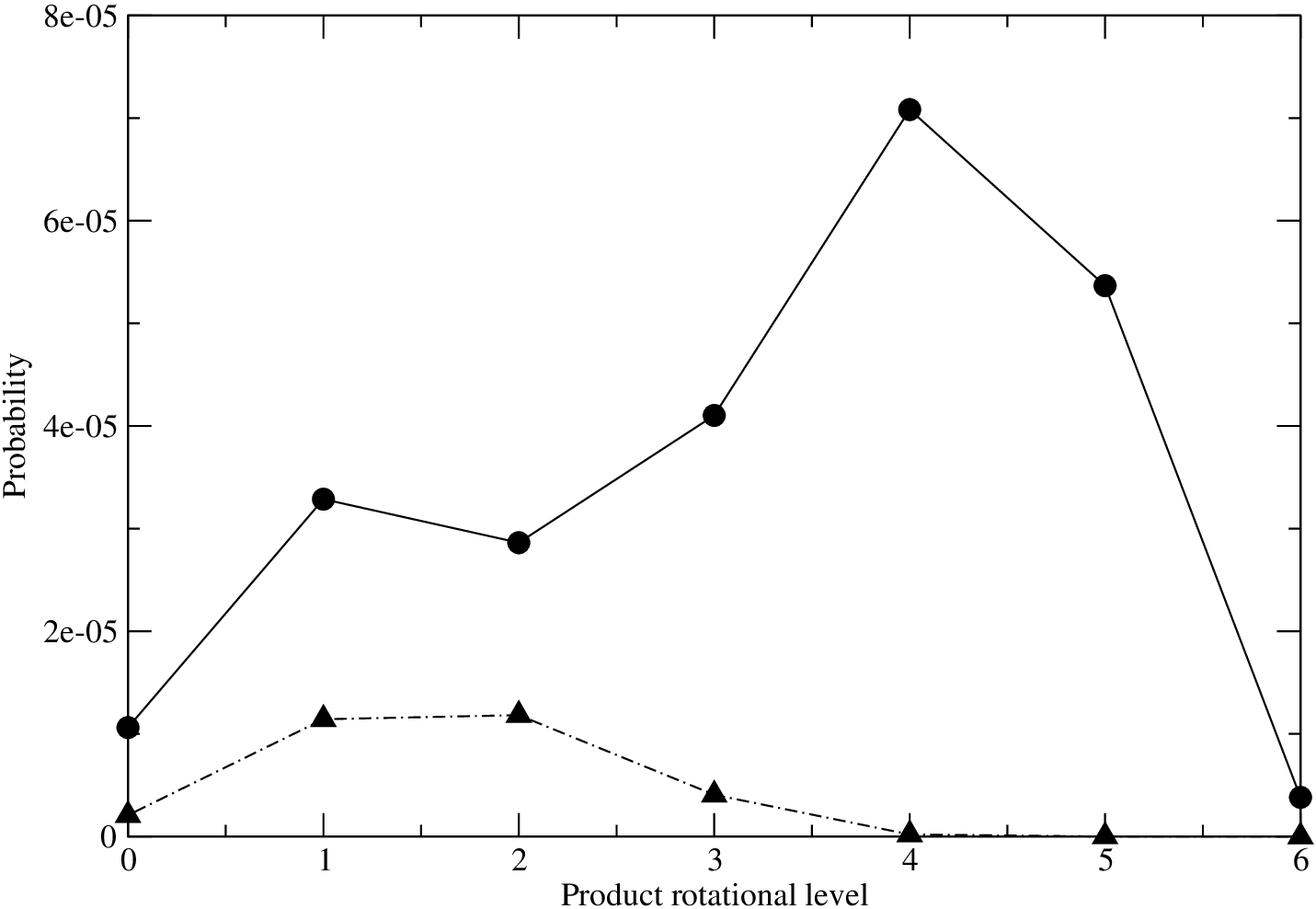}}
\end{center}
\caption{\label{fig:5} Comparison of the state-to-state reaction probabilities 
 for $\mbox{H}_2(v'=0,j')$ (solid curve) and $\mbox{HD}(v'=0,j')$ (dot-dashed curve) 
 formation in $\mbox{H}+\mbox{HCl}(v=1,j=0)$ and $\mbox{H}+\mbox{DCl}(v=1,j=0)$ 
 collisions, respectively. The probability is represented as a function 
 of the product rotational number $j'$ for a fixed incident kinetic energy of 
 $10^{-5}~\mbox{eV}$.
 }
\end{figure}

State-to-state reaction probabilities as a function of the product rotational quantum 
number $j'$ are represented in Figure \ref{fig:5}~ for the ground vibrational state of 
the $\mbox{H}_2$ and $\mbox{HD}$ fragments in $\mbox{H}+\mbox{HCl}(v=1,j=0)$ and 
$\mbox{H}+\mbox{DCl}(v=1,j=0)$ collisions, respectively. For a fixed incident kinetic 
energy of $10^{-5}~\mbox{eV}$, 7 rotational levels are energetically 
accessible in the diatomic products of both reactions. High-$j$ channels of 
$\mbox{H}_2$ are preferentially populated even though lower-lying rotational states 
are open. The probability for $\mbox{H}_2$ formation in its ground vibrational state 
peaks at $j'=4$, corresponding to an exoergicity of 
$0.2379~\mbox{eV}=5.487~\mbox{kcal/mol}$ for the 
abstraction reaction. In the case of HD$(v'=0)$ formation, the maximum energy released by 
this reaction is $0.2354~\mbox{eV}=5.429~\mbox{kcal/mol}$, for $j'=2$, with a significantly 
reduced probability than for the $\mbox{H}_2$ case.        

\begin{figure} 
\begin{center}
\resizebox{0.49\textwidth}{!}{\includegraphics{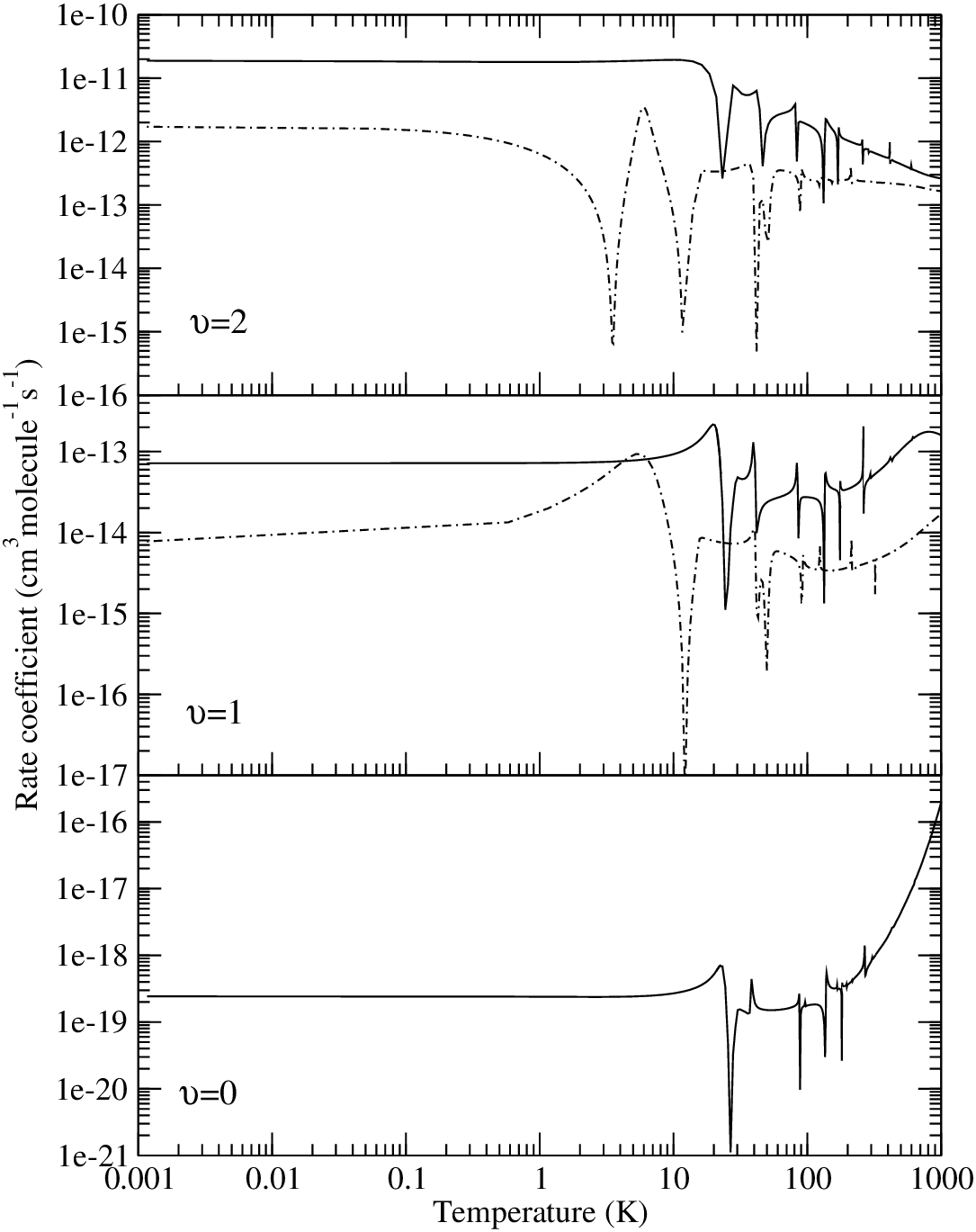}}
\end{center}
\caption{\label{fig:6} Reaction rate coefficients for H$_2$ 
 and HD formation for the $\mbox{H}+\mbox{HCl}(v,j=0)$ (solid curve) 
 and $\mbox{H}+\mbox{DCl}(v,j=0)$ (dot-dashed curve) reactions, for $v=0-2$, as 
 a function of the temperature.}
\end{figure}

Figure \ref{fig:6} displays the reaction rate coefficients 
defined as the product of the cross section and the relative velocity for H$_2$ 
and HD formation for the $\mbox{H}+\mbox{HCl}(v,j=0)$ 
and $\mbox{H}+\mbox{DCl}(v,j=0)$ reactions, respectively, for $v=0-2$, as 
a function of the temperature. 
The Wigner regime, for which the rate coefficients become constant, is attained 
for temperatures below $1~\mbox{K}$ for H$_2$ formation, 
and below $10^{-3}~\mbox{K}$ for the HD product. For $v=1$ and 2, the rate 
coefficients for H$_2$ production is an order of magnitude larger than for HD at 
cold and ultracold temperatures. However, for $v=1$ the rate coefficient 
for HD formation becomes slightly larger than for H$_2$ in the tunneling region of the 
$\mbox{H}+\mbox{DCl}$ reaction, i.e. around $T=5~\mbox{K}$.    
In the zero-temperature limit, the rate coefficients calculated for H$_2$ and HD 
formation, respectively, are $1.9\times 10^{-11}~\mbox{cm}^3~\mbox{s}^{-1}$ and 
$1.7\times 10^{-12}~\mbox{cm}^3~\mbox{s}^{-1}$ for $v=2$, 
$7.2\times 10^{-14}~\mbox{cm}^3~\mbox{s}^{-1}$ and 
$7.8\times 10^{-15}~\mbox{cm}^3~\mbox{s}^{-1}$ for $v=1$, 
and $2.4\times 10^{-19}~\mbox{cm}^3~\mbox{s}^{-1}$ 
for H$_2$ with $v=0$. 

\section{Conclusion}  \label{sec:conclu}

State-to-state and initial-state-selected cross sections have been 
calculated using fully quantum mechanical techniques for both reactive 
and non-reactive channels of the $\mbox{H}+\mbox{HCl}(v,j=0)$ and 
$\mbox{H}+\mbox{DCl}(v,j=0)$ collisions, for $v=0-2$. Resonance 
structures due to quasibound states 
of the H$\cdots$HCl and H$\cdots$DCl van der Waals wells in the entrance valley 
appear in energy dependence of the cross sections. Our results also indicate that H$_2$ 
formation is the predominant process of $\mbox{H}+\mbox{HCl}$ collisions at 
cold and ultracold temperatures, while for $\mbox{H}+\mbox{DCl}$ non-reactive 
scattering is more favorable in this regime. We find that, for both 
$\mbox{H}+\mbox{HCl}$ and $\mbox{H}+\mbox{DCl}$ collisions, vibrational 
excitation dramatically enhances the zero-temperature limiting value of the 
rate coefficients. The effect of vibrational excitation is found to be comparable for 
reactive and nonreactive channels in the ultracold limit.

\begin{acknowledgement}

This work was supported by NSF grant PHYS-0245019, the Research Corporation and by the 
United States-Israel Binational Science Foundation.

\end{acknowledgement}



\end{document}